\documentclass[number,sort&compress]{elsarticle}

\usepackage{aas_macros}

\usepackage{amsmath,amssymb}

\usepackage{threeparttable}

\usepackage{hyperref}

\DeclareMathOperator{\atantwo}{atan2}

\begin{document}

\begin{frontmatter}

\title{A convolutional neural network approach for reconstructing polarization information of photoelectric X-ray polarimeters}

\author[a]{Takao~Kitaguchi}
\ead{takao.kitaguchi@riken.jp}

\author[b,c]{Kevin~Black}
\author[d]{Teruaki~Enoto}
\author[a]{Asami~Hayato}
\author[b]{Joanne~E.~Hill}
\author[e]{Wataru~B.~Iwakiri}
\author[f]{Philip~Kaaret}
\author[g,h,i]{Tsunefumi~Mizuno}
\author[a]{Toru~Tamagawa}

\address[a]{RIKEN Nishina Center, 2-1 Hirosawa, Wako, Saitama 351-0198, Japan}
\address[b]{NASA Goddard Space Flight Center, Greenbelt, MD 20771, USA}
\address[c]{Rock Creek Scientific, 1400 East-West Hwy, Silver Spring, MD, 20910, USA}
\address[d]{The Hakubi Center for Advanced Research, Kyoto University, Kyoto 606-8302, Japan}
\address[e]{Department of Physics, Faculty of Science and Engineering, Chuo University, 1-13-27 Kasuga, Bunkyo-ku, Tokyo 112-8551, Japan}
\address[f]{University of Iowa, Iowa City, IA, 52242, USA}
\address[g]{Department of Physical Science, Hiroshima University, 1-3-1 Kagamiyama, Higashi-Hiroshima, Hiroshima 739-8526, Japan}
\address[h]{Core Research for Energetic Universe, Hiroshima University, 1-3-1, Kagamiyama, Higashi-Hiroshima, Hiroshima 739-8526, Japan}
\address[i]{Hiroshima Astrophysical Science Center, Hiroshima University, 1-3-1 Kagamiyama, Higashi-Hiroshima, Hiroshima 739-8526, Japan}

\begin{abstract}
This paper presents a data processing algorithm with machine learning for polarization extraction and event selection applied to photoelectron track images taken with X-ray polarimeters. The method uses a convolutional neural network (CNN) classification to predict the azimuthal angle and 2-D position of the initial photoelectron emission from a 2-D track image projected along the X-ray incident direction. Two CNN models are demonstrated with data sets generated by a Monte Carlo simulation: one has a commonly used loss function calculated by the cross entropy and the other has an additional loss term to penalize nonuniformity for an unpolarized modulation curve based on the $H$-test, which is used for periodic signal search in X-ray/$\gamma$-ray astronomy. The modulation curve calculated by the former model with unpolarized data has several irregular features, which can be canceled out by unfolding the angular response or simulating the detector rotation. On the other hand, the latter model can predict a flat modulation curve with a residual systematic modulation down to $\lesssim1$\%. Both models show almost the same modulation factors and position accuracy of less than 2~pixel (or 240~$\mu$m) for all four test energies of 2.7, 4.5, 6.4, and 8.0~keV. In addition, event selection is performed based on probabilities from the CNN to maximize the polarization sensitivity considering a trade-off between the modulation factor and signal acceptance. The developed method with machine learning improves the polarization sensitivity by 10--20\%, compared to that determined with the image moment method developed previously.
\end{abstract}

\begin{keyword}
X-ray, photoelectric polarimeter, tracking algorithm, machine learning, convolutional neural network
\end{keyword}

\end{frontmatter}


\section{Introduction}
\label{sect:intro}  

Recent developments in micropattern gas detectors enable us to track an electron with an energy down to 1~keV. These detectors can be used as X-ray polarimeters by imaging a photoelectron track induced by an X-ray and extracting the polarization from its initial direction. Since the photoelectric effect is the dominant interaction for X-rays, the micropattern gas polarimeter combined with an X-ray focusing optics is expected to achieve a much higher sensitivity than the Bragg diffraction polarimeter \cite{1976ApJ...208L.125W} that is the only satellite-borne polarimeter flown in space so far in the soft X-ray energy band less than 10~keV.

The differential cross section of K-shell photoelectrons in the non-relativistic region is proportional to $\cos^{2}(\phi)$, where $\phi$ is the azimuthal angle with respect to the X-ray electric vector \cite{1959PhRv..113..514G}. The modulation amplitude, $a$, which is defined as the ratio of sinusoidal amplitude to unmodulated offset derived from the photoelectron angular distribution (or so-called the modulation curve), is intrinsically 100\% for perfectly linearly polarized X-rays. However, the observed modulation amplitude declines due to imperfect angular reconstruction of photoelectron track images which are curved by Coulomb scattering and are blurred by electron diffusion (see Fig.~1 in Ref.~\citenum{2018NIMPA.880..188K}). As accurate an angular reconstruction method as possible is preferred to achieve the modulation amplitude close to 100\%. The modulation amplitude for perfectly polarized X-rays is called the modulation factor designated as $\mu$. The observed modulation amplitude is then give by $a = a_{p} \mu$, where $a_{p}$ is the polarization degree of a source.

So far, two major reconstruction methods have been demonstrated: one is with image moments (e.g. \cite{2003SPIE.4843..383B,2018NIMPA.880..188K}) and the other is based on the shortest path problem in graph theory \cite{2017NIMPA.858...62L}. When applying the two methods to track images from the polarimeter we have developed \cite{2007NIMPA.581..755B,2014SPIE.9144E..1NH}, the results were found to be similar \cite{2018NIMPA.880..188K}. Since both methods eventually use image moments to reconstruct the initial photoelectron emission direction, it may show a limitation of the image moments method, which is representative measure but with the loss of some information. Furthermore, image pre-processing to calculate image moments loses a part of pixel information mainly due to image thresholding. Therefore, a further improvement of the polarimeter sensitivity requires a novel reconstruction algorithm which can deal with all the pixels of a track image.

An alternative approach to determine the initial direction and position of a photoelectron track is to learn the image features automatically with a machine. Machine learning is a rapidly evolving field of computer science and can provide many different approaches to make data-driven prediction by building a model from sample inputs. Among them, convolutional neural networks (CNNs) have generally shown an excellent performance in image recognition. For example, CNNs achieve a low misclassification error rate for the Modified National Institute of Standards and Technology (MNIST) database \cite{Lecun98} on handwritten digits (see its official site\footnote{http://yann.lecun.com/exdb/mnist/} for recent results).

This paper describes polarization extraction and track selection based on a CNN. Section~\ref{sect:cnn} briefly explains the CNN model developed for this work. We describe results derived from a commonly used loss function combined with the CNN model in Section~\ref{sect:bas} and a modified one for polarimetry in Section~\ref{sect:mod}. In addition, track selection with CNN-derived products is demonstrated in Section~\ref{sect:sel}. Lastly, we conclude the study in Section~\ref{sect:con}. Throughout this paper, all errors are given at the $1\sigma$ confidence level unless otherwise stated.

\section{Convolutional neural network}
\label{sect:cnn}

\subsection{Overview}
\label{sect:ov}

The main goal in this paper is to extract polarization information from 2-D images of photoelectron tracks with a CNN. An artificial neural network is made up of several layers, each of which contains a number of nodes. A value of each node is computed as a bias plus a linear combination of (all or a part of) nodes in the previous layer, followed by a nonlinear transformation with a so-called activation function. The weights of the linear combination and biases are free parameters and are optimized through training with an example data set of input-output pairs by fitting outputs predicted from the network to corresponding inputs; that is supervised learning.

Supervised learning is mainly categorized into two methods: regression computing a continuous value and classification predicting a discrete label to which an input belongs. In classification, the output layer contains votes (or probabilities which are normalized votes) of each label and selects the largest vote (or highest probability) to predict the label. In this work, multi-class classification is employed because the highest probability is efficient in selecting good tracks for polarization measurements, as described in detail later. In addition, classification is generally more robust to outliers than standard regression with the least squares approach.

A CNN is a type of feed-forward artificial neural network where data pass only in the forward direction from an input layer, through multiple hidden layers, and finally to an output layer, with no loops or cycles. The hidden layers of a CNN contain a combination of three types of layers: convolutional, pooling, and fully connected layers. The convolutional layer is a key component of a CNN to automatically extract image features by performing a convolution operation on an input with a weight matrix called a filter or kernel. The operation is repeated by shifting the filter over an input image with a step size called a stride. The pooling layer performs nonlinear downsampling to reduce the size of the convolved images while maintaining their features. The fully connected layer connects every node in one layer to every node in another layer, or in mathematical words, multiplies the input nodes by a weight matrix and then adds a bias vector. Following a single- or multi-set of the convolutional and pooling layers, the fully connected layers take image features from the last pooling layer and classify them into various classes (or regress from them to a numerical value).

\subsection{Network model}
\label{sect:nm}

\begin{figure}
 \centering
 \includegraphics[width=6.5cm]{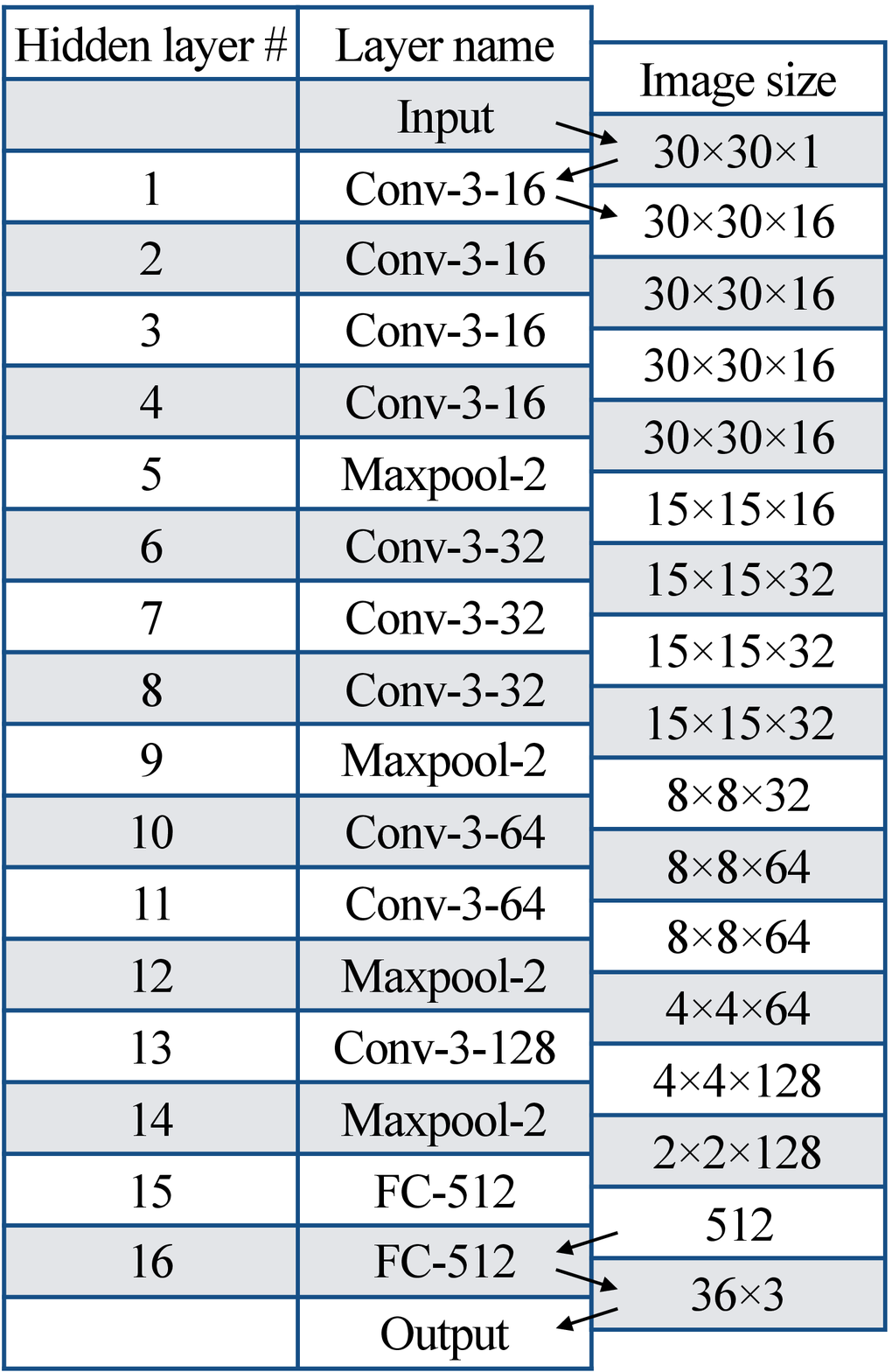}
 \caption
 {\label{fig:cnn}
 Schematic view of the CNN model designed for polarization reconstruction. The convolutional layer parameters are denoted as Conv-(square filter size)-(number of channels). Maxpool-2 represents the max pooling layer with a $2\times2$ filter. FC-512 indicates the fully connected layer with 512 nodes.}
\end{figure}

Figure~\ref{fig:cnn} illustrates a schematic view of the CNN classification model that we designed for X-ray polarimetry. The number of network parameters is 0.74 million. The input layer consists of 900 nodes corresponding to the $30\times30$ pixels of a photoelectron track image. Since polarization measurements requires at least the initial direction of photoelectron emission, the output layer should have angular predictions. In addition, for imaging capability, the output layer is preferred to include predictions for the initial photoelectron position or the X-ray interaction point. Therefore, the output layer is designed to simultaneously predict the three values: the initial azimuthal angle and 2-D position of photoelectron emission. Each prediction is represented by a probability histogram with 36 equal-width bins; for example, the probability distribution of the predicted angle ranging from -90 to 90~deg\footnote{Since the differential cross section of photoelectron emission is represented with a bimodal sinusoid, a twofold angular distribution is sufficient enough to determine polarization.} is divided by 36 bins each having a width of 5~deg. Thus, the output layer has $36~\mathrm{bins} \times 3~\mathrm{predictions} = 108$ nodes in total. Although the number of probability bins is arbitrary, its optimization is out of scope for this work.

The hidden layers are constructed based on the Visual Geometry Group (VGG) model \citep{2014arXiv1409.1556S}, and are comprised of the convolution layers with the filer size $3\times3$ and stride 1 and the maximum pooling layers with the filter size $2\times2$ and stride 2, and the fully-connected layers. The VGG developers demonstrated that multiple continuous convolution layers with a size of $3\times3$ is more effective than a single convolution layer with a wider kernel. All the layers except for the pooling layers and last hidden layer utilize the rectified linear unit (ReLU \citep{Nair10}) function as the activation function. The last fully-connected layer uses the softmax function to compute each of the three probability distributions.

For each event, the actual initial direction and 2-D position of the photoelectrons in the training data are separately converted into a one-hot vector where an element to which the answer belongs is 1 and all the other 35 elements are 0 to correspond with the bin size of the outputted probability histogram. In order to evaluate a difference between each of the three probability distributions and the one-hot answer vector, the cross entropy, $H$, between them is computed. The sum of the three cross entropies is used as the loss function to evaluate the fitting (or supervise the machine learning) and is minimized with an optimization algorithm. In addition, L2 regularization, which is the sum of the square of the weights, is added to $H$ to suppress overfitting. The loss function, $L$, of the network model is given by:
\begin{equation}\label{eq:LH2}
 L = \sum H + \lambda_{2} \sum w^{2},
\end{equation}
where $w$ is a weight and $\lambda_{2}$ is a hyperparameter, which is fixed at a constant value set before the training process, for L2 regularization to control trade-off between overfitting and training efficiency.

In order to train the CNN model (or optimize its weights and biases), we employ the Adam optimizer \citep{2014arXiv1412.6980K}, a commonly used type of stochastic gradient descent. In addition, mini-batch training with a size of 360 images is performed to efficiently find the minimum of $L$.

In general, a deeper feed-forward network can lead to better performance (e.g. \citep{Montufar:2014:NLR:2969033.2969153}). Although another set of convolution and pooling layers, or fully-connected layers were added to the network in Fig.~\ref{fig:cnn}, the result did not significantly change. In addition, wider networks were found to show no improvement.

\subsection{Input data: photoelectron track images}
\label{sect:inp}

Supervised machine learning requires a training data set containing input-output pairs. In this case, the training set is a set of photoelectron track images with initial direction and 2-D position. The photoelectric polarimeter that we have developed can make a photoelectron track with $30\times30$ pixels of a size of $121\times121~\mu$m \citep{2014SPIE.9144E..1NH}. Although, in actual experiment, we can easily regulate the X-ray polarization direction, we cannot control the photoelectron direction which is randomized according to the differential cross section, $\cos^{2}(\phi)$. Therefore, a Monte Carlo simulation was performed to generate the training data.

The simulator designed and developed for the time projection chamber (TPC) polarimeter \citep{2014SPIE.9144E..4LK} can reproduce the passage of X-rays and electrons through dimethyl ether gas at a pressure of 190~Torr, and generate photoelectron track images with the same dimensions as physical ones. It accounts for detector effects such as image blurring due to diffusion of drifting electrons to a readout electrode, electron amplification by a gas electron multiplier (GEM), and image elongation due to signal shaping with a time constant of 50~ns. The readout scheme, which is triggered by the GEM cathode signal, is also simulated to take a 2-D track image with the TPC technique, where signal charges are read out by 1-D strip electrodes with continuous analog-to-digital converter (ADC) sampling at a frequency of 20~MHz. The readout image size is $30\times30$ pixels, each having an ADC value.

\begin{figure}
 \begin{center}
  \includegraphics[width=7.0cm]{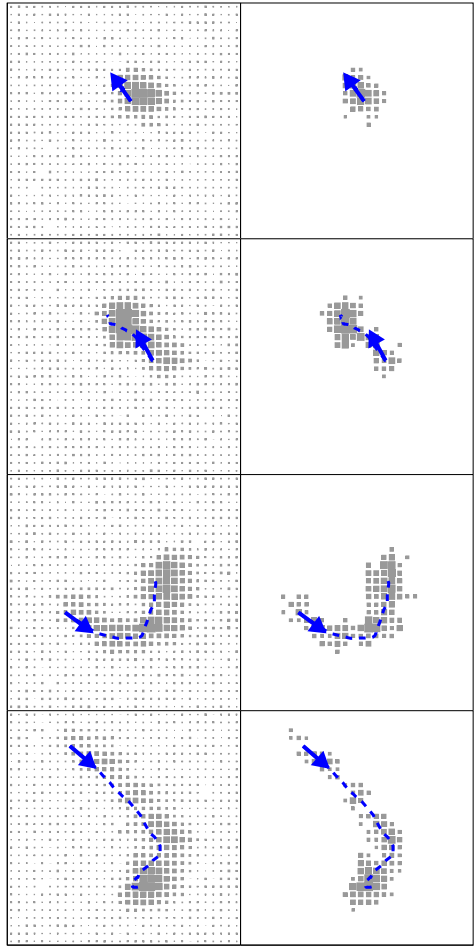}
 \end{center}
 \caption
 { \label{fig:imgs}
 Typical photoelectron track images generated by the Monte Carlo simulation. The left and right panels show raw and offline processed images, respectively. From top to bottom, the incident X-ray energy is 2.7, 4.5, 6.4, and 8.0~keV. The vertical position is determined with the readout strip position, while the horizontal one is measured with the time variation of charge amounts induced by electrons drifting to the corresponding strips. The gray box size in each pixel is proportional to the charge amount. The direction and start position of the blue arrows indicate the initial angle and point of photoelectron emission, respectively. The blue dashed curves show photoelectron trajectory. Compared to the processed images, the corresponding raw images are elongated on the vertical axis due to signal shaping with a time constant of 50~ns or 1~pixel size.}
\end{figure}

The simulated images can be offline processed in the same manner as actual images as described in \S~3 of Ref.~\citenum{2016NIMPA.838...89I}. The offline image processing mainly includes deconvolution with the signal shaping response and then image thresholding to perform the conventional angular reconstruction methods. Example raw and corresponding processed images are shown in Fig.~\ref{fig:imgs}. The modulation factors derived from the processed images with the image moment method can be reproduced with an absolute error of $\pm3$\% (or a relative error of $\sim5$\%) in the 2.7--8.0~keV energy range, where the polarimeter performance was evaluated in detail with a synchrotron X-ray facility \citep{2016NIMPA.838...89I}.

As input data to the network, we used the raw images elongated on the vertical axis in Fig.~\ref{fig:imgs} by the signal shaping process rather than the offline processed images which were utilized for the angular reconstruction methods so far reported. This is because the image processing loses some of image information mainly by image thresholding which sets a charge amount to zero if it is below a set threshold. Furthermore, deconvolution of the signal shaping response makes the signal-to-noise ratio worse. Since a CNN is known as a robust model that needs relatively little image pre-processing, only the linear normalization is applied to our data sets so that the minimum and maximum values among all the pixels in all the images are 0 and 1, respectively.

The actual initial direction and 2-D position of photoelectron emission to supervise the network are individually discretized and grouped into 36 classes corresponding to the outputs of the CNN model. For example, the class $i$ ($i=0,1,\cdots,35$) for the photoelectron direction constitutes angles ranging from $5i-90$ to $5i-85$~deg. The emission position is originally calculated as a relative coordinate value in the detector coordinate system and is converted into a float value in the image pixel coordinate with a $30 \times 30$ pixel size. Then, the emission angle ranging from $30i/36$ to $30(i+1)/36$~pixel is categorized as the class $i$. When the emission position is located outside of the image frame, the position class is floored/ceiled to 0/35.

Classification with imbalanced training data may result in poor performance (e.g. \cite{5128907}). In order to avoid the imbalanced learning problem, the training data containing balanced angular classes were generated with the simulator by irradiation of unpolarized rather than polarized X-rays. In addition, a wide variety of raw track images were created by setting the primary X-ray energy ranging from 0.5 to 12~keV and irradiation distance from (or electron drift length to) the readout electrode ranging from 0.45 to 1.55~cm, the increase of each parameter makes a photoelectron track longer and more blurred, respectively. The X-ray momentum direction was fixed parallel to the optical axis of the polarimeter, while the X-ray polarization direction was set to be randomly distributed perpendicular to its momentum direction. The X-ray energy is randomly sampled between 0.5 and 12~keV with weighting by inverse of detection efficiency to make the resulting spectrum flatten. In this way, we generated 3,600,000 and 360,000 images for unpolarized X-rays as training and validation data sets, respectively.

In addition to the training and validation data sets, unpolarized and perfectly polarized data sets each containing monochromatic (2.7, 4.5, 6.4, and 8.0~keV, separately) X-rays were generated with the same simulator to test the network model after training. The irradiation position for the test is limited to between 0.6 and 1.0~cm unlike the training and verification data sets. In addition, for the polarized data, the X-ray polarization direction is inclined at $\phi = -45$~deg in the image coordinate defined in Fig.~\ref{fig:imgs}. These are because the polarized test data set is a simulation to reproduce those actually observed with an X-ray synchrotron radiation facility \cite{2016NIMPA.838...89I}. Each test data set at monochromatic energy contains 360,000 track images. Only for comparison purposes with the machine learning (ML) approach, the test raw images are processed in the conventional way separately with the image moment (IM) and graph-based (GR) methods to calculate the modulation factor.

\section{Experiments}
\label{sect:exp}

\subsection{Basic model}
\label{sect:bas}

We programmed the network model shown in Fig.~\ref{fig:cnn} with Google TensorFlow \citep{2016arXiv160304467A} version 1.8.0 and trained it with a NVIDIA GeForce GTX 1080 graphics processing unit (GPU). At the beginning of training, all biases were set to zero, while all weights were initialized to uniform random values dedicated to the ReLU activation function according to Ref.~\citenum{DBLP:journals/corr/HeZR015}. The hyperparameter $\lambda_{2}$ in Eq.~(\ref{eq:LH2}) should be low to increase training efficiency without overfitting. It was optimized with the grid search method ranging from 0.0001 to 0.0010 with step 0.0001 and was set to 0.0007 because overfitting appears if $\lambda_{2} \leq 0.0006$.

\begin{figure}
 \begin{center}
  \includegraphics[width=7.0cm]{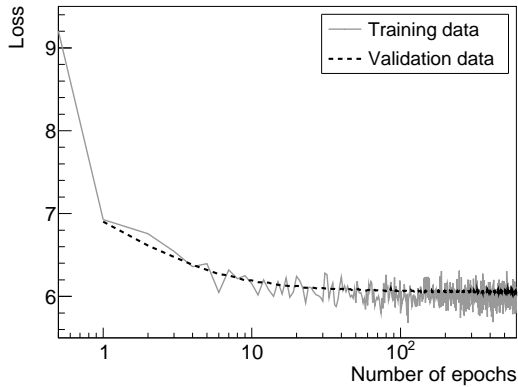}
 \end{center}
 \caption
 { \label{fig:lc}
 Learning curve as a function of the number of epochs for the (gray solid) training and (black dashed) validation images.}
\end{figure}

We used the 3,600,000 images for training and performed stochastic mini-batch training with a size of 360 images to efficiently find the minimum of the loss function of Eq.~(\ref{eq:LH2}) and suppress overfitting. After every epoch, which means an iteration over all images in the training data set, we evaluated the cross entropy (or goodness of fitting) with the 360,000 validation images. A learning curve shown in Fig.~\ref{fig:lc} shows that training is successful without overfitting -- the loss of both training and validation data sets decreases and approaches to a constant value as training progresses. The total computation time for 600 epochs with the GPU is approximately 45 hours.

\begin{figure}
 \begin{center}
  \includegraphics[width=7.0cm]{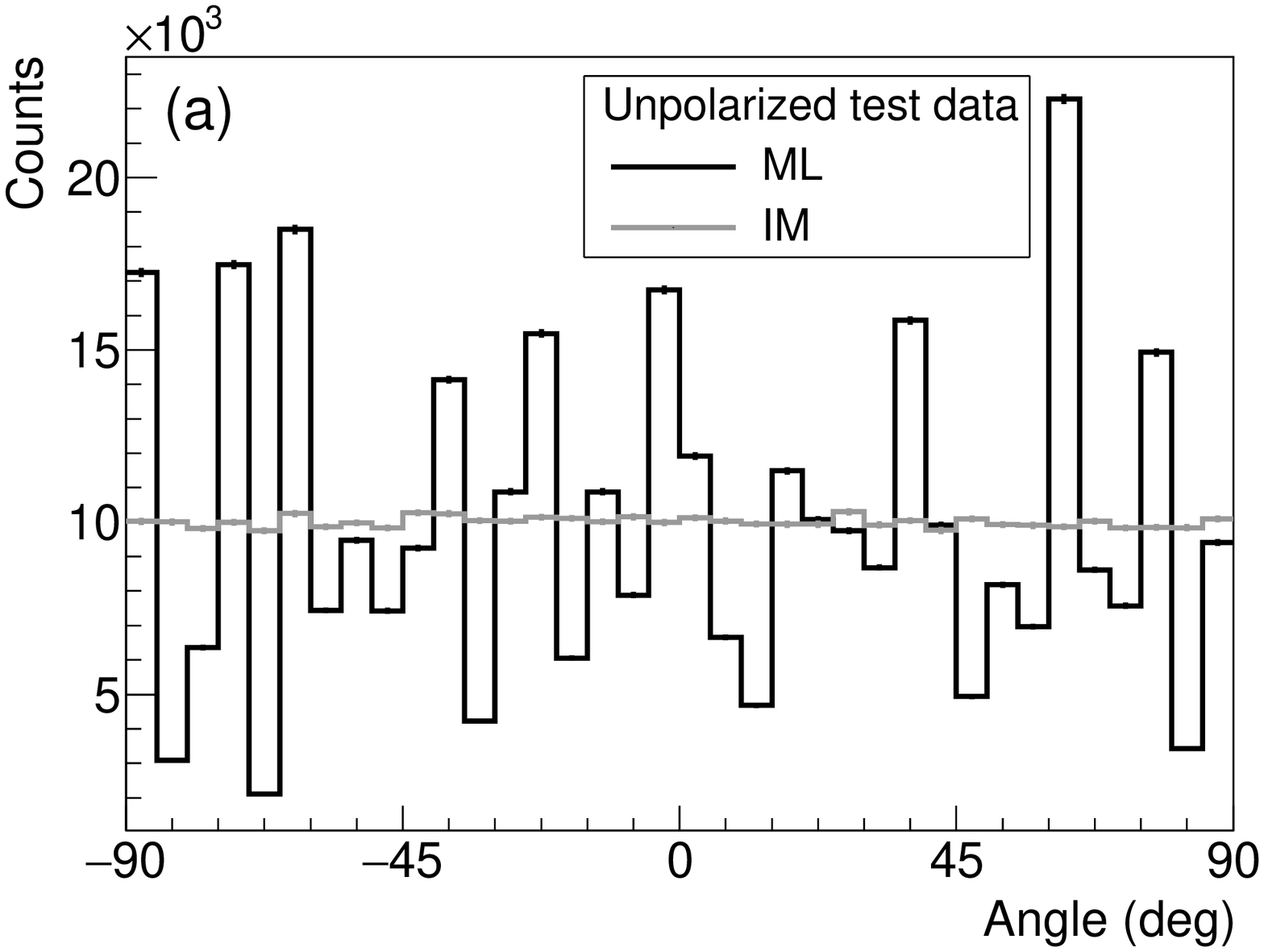}
  \includegraphics[width=7.0cm]{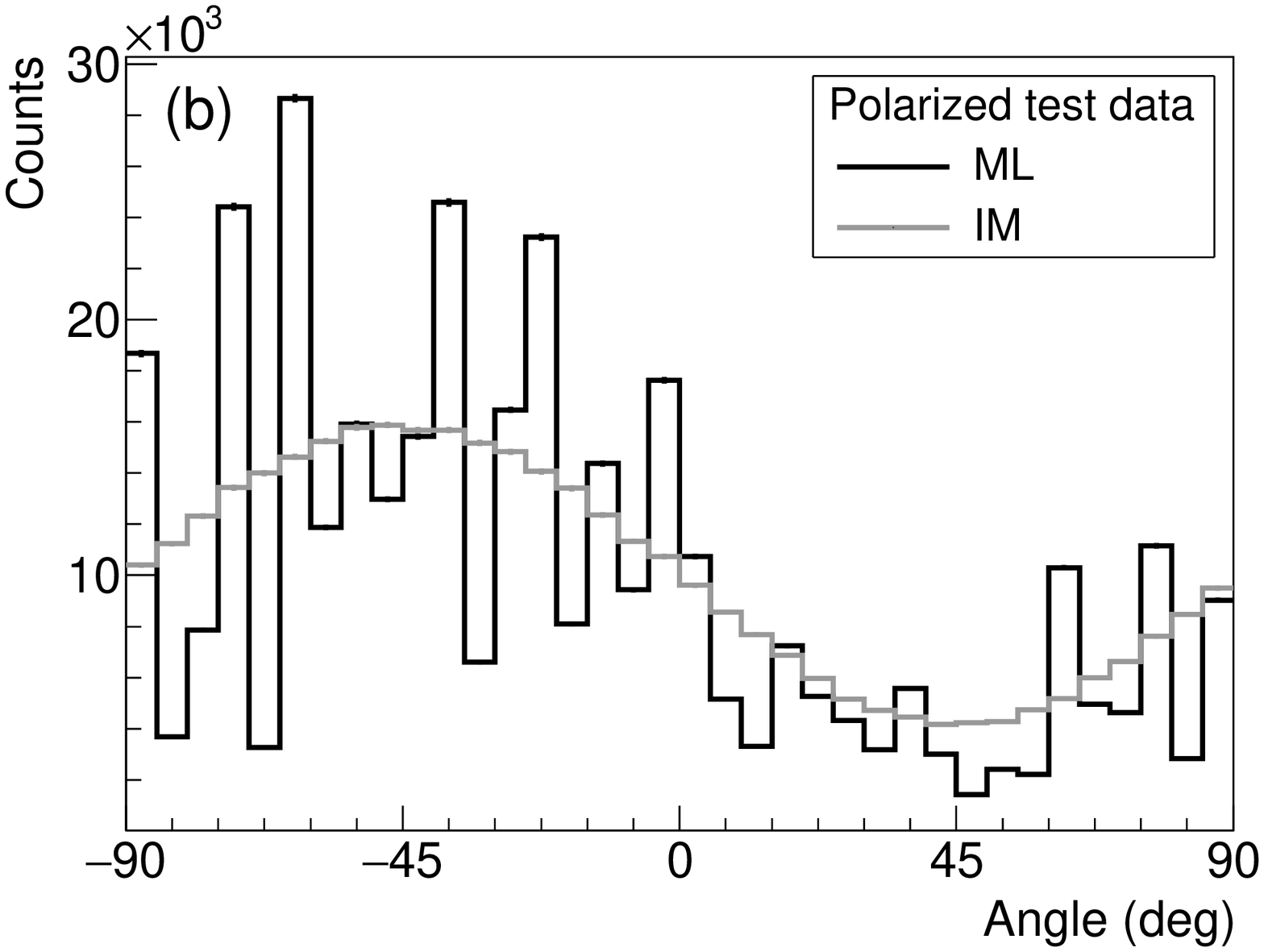}
 \end{center}
 \caption
 { \label{fig:CompMC}
 Modulation curves created with angles predicted by machine learning (ML, black) and reconstructed by image moments (IM, gray) for comparison. Panel~(a) represents modulation curves for the unpolarized test data with an X-ray energy of 8.0~keV. Panel~(b) is the same as (a), but for the perfectly linearly polarized test data with an instrinsic polarization angle of -45~deg.}
\end{figure}

During training, we evaluated the network model after every epoch by inputting the validation data set and stored the one with the minimum of the loss. Figure~\ref{fig:CompMC}~(a) shows the modulation curve created by predicted angles from the best (or the lowest loss) model, which were saved at the epoch of 460, by inputting the simulated test data sets of unpolarized 8.0~keV X-rays. For comparison, the modulation curve calculated with the IM method is also shown in Fig~\ref{fig:CompMC}~(a). Although a modulation curve for unpolarized X-rays is preferred to be directionally uniform to simplify the polarimetry analysis (e.g. \cite{2017ApJ...838...72S}), the unpolarized modulation curve generated by the ML model shows an obvious residual modulation with several irregular features. On the other hand, the IM modulation curve is flat with a residual modulation amplitude, $a$, of $0.9 \pm 0.2$\%, which is derived by fitting it with a sinusoidal model of $C\{a\cos[2(\phi-\phi_{0})]+1\}$, where $C$ is a constant factor or unmodulated offset and $\phi_{0}$ is the polarization angle. In a similar way, the polarized modulation curve shown in Fig~\ref{fig:CompMC}~(b) is not sinusoidal, while that obtained with IM shows a unimodal sinusoidal curve with $a = \mu = 58.1 \pm 0.2$\% and $\phi_{0} = -44.7 \pm 0.1$~deg.

\begin{figure}
 \begin{center}
  \includegraphics[width=7.0cm]{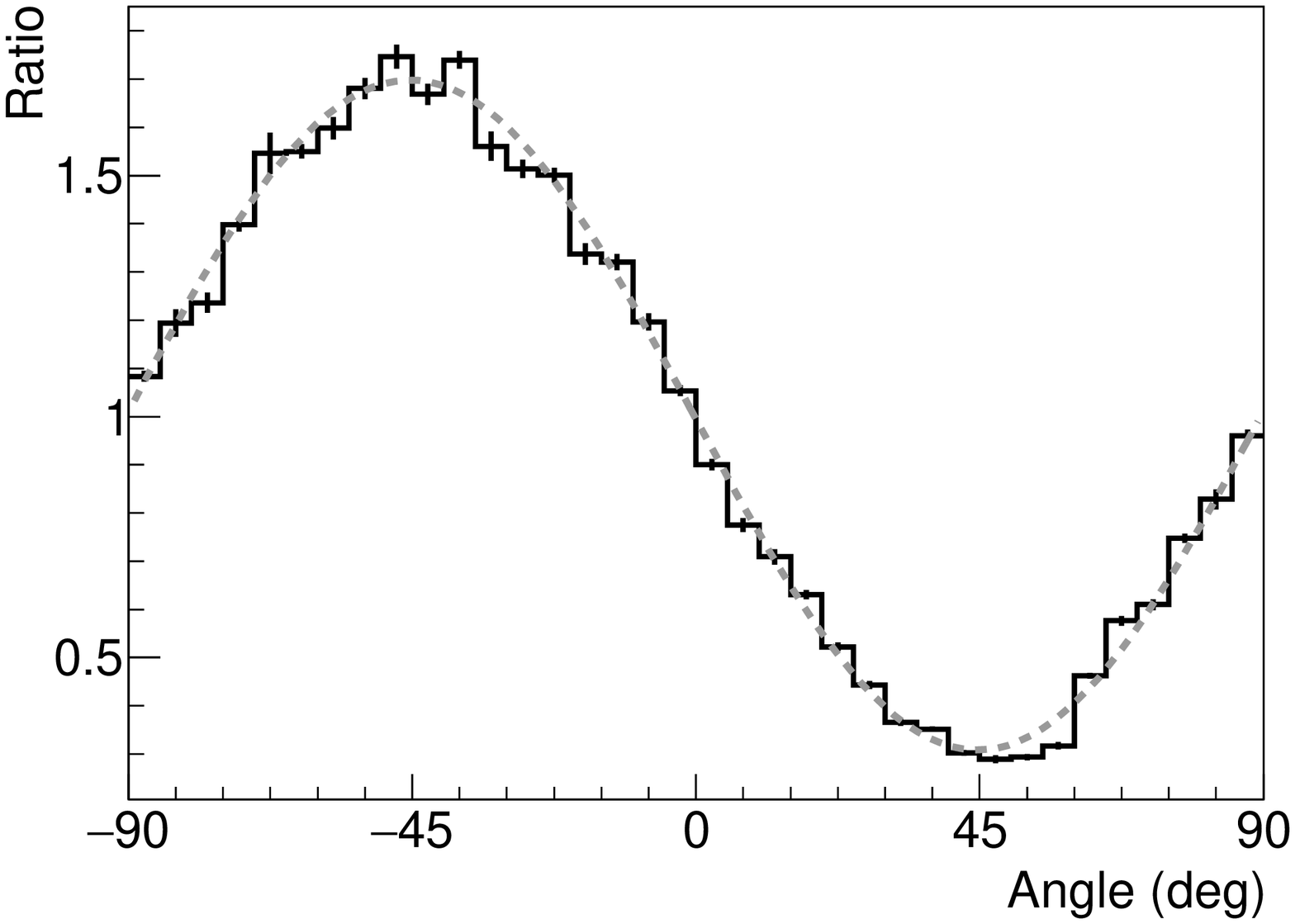}
 \end{center}
 \caption
 { \label{fig:UnfoldMC}
 Unfolded modulation curve for 8.0~keV X-rays. The black solid histogram is created by dividing the polarized curve by the unpolarized one shown in black in Fig.~\ref{fig:CompMC}. The gray dashed curve is the best-fit sinusoidal model with a modulation factor of 69.2\% and a phase of -45.3~deg.}
\end{figure}

Two methods are suggested to cancel out the systematic irregular modulation and extract polarization information from the modulation curves. The first method is unfolding the polarized modulation curve with the unpolarized one. The simplest method of an unfolding is dividing the polarized curve by the unpolarized one\footnote{Similar method to the so-called Crab ratio to unfold a spectral response function \cite{2000AdSpR..25..399D}, which divides a source spectrum by the Crab spectrum observed with the same detector.}. The divided modulation curve in Fig.~\ref{fig:UnfoldMC} shows a unimodal sinusoidal shape with $\mu = 69.2\pm0.2$\% and $\phi_{0} = -45.3\pm0.1$~deg.
A $\chi^{2} $ value of the fit is 279.8 with 33 degrees of freedom (dof), indicating that there still remain small residuals from the best-fit sinusoidal model because the division method is only an approximate unfolding. To be more accurate, an unfolding with a 2-d response function of the X-ray polarization angle vs. the predicted one is needed.

The second method needs uniform rotation of polarimeter or satellite with respect to the optical axis. In fact, the PRAXyS X-ray polarimetry mission had a plan to rotate its satellite with rotations per minute of 10 to reduce a residual modulation \cite{2016SPIE.9905E..16J}. The rotation is simulated with the same unpolarized data that the first method is applied to, by shifting predicted angles so that the X-ray polarization angle, which is initially set to a random value, goes to -45~deg in the modulation curve coordinate. The resulting modulation curve is similar to Fig.~\ref{fig:UnfoldMC}~(b), but has smaller residuals from the best-fit sinusoidal model with $\chi^{2}/\mathrm{dof} = 33.2/33$. The modulation factor and polarization degree are respectively $69.0\pm0.2$\% and $-45.1\pm0.1$~deg, which are consistent with those determined with the first method.

Table~\ref{tab:summary} summarizes the fitting results for all the monoenergetic test data sets with the different angular reconstruction methods including the conventional IM and GR methods and more sophisticated ones described later. It shows that machine learning can improve the modulation factors by $\sim10$\% and $\sim20$\% in the lower and higher energy ranges, respectively.

\begin{threeparttable}
 \footnotesize
 \centering
 \caption
 { \label{tab:summary}
 Best-fit parameters of the sinusoidal curve model to modulation curves generated with the different angular reconstruction methods.}
 \begin{tabular}{l l l l l l l l l}
  \hline 
  \hline 
  Energy & Method   & & \multicolumn{2}{c}{Unpolarized} & & \multicolumn{3}{c}{Polarized\tnote{a}} \\ \cline{4-5} \cline{7-9}
  \multicolumn{1}{c}{(keV)} &          & & \multicolumn{1}{c}{$\mu$ (\%)} & \multicolumn{1}{c}{$\chi^{2}$~\tnote{b}} & & \multicolumn{1}{c}{$\mu$ (\%)} & \multicolumn{1}{c}{$\phi_{0}$ (deg)} & \multicolumn{1}{c}{$\chi^{2}$~\tnote{b}} \\
  \hline 
  2.7    & IM\tnote{c}       & & $0.4 \pm 0.2$ & 23.1 & & $23.4 \pm 0.2$ & $-45.1 \pm 0.3$ & 41.4 \\
         & GR\tnote{d}       & & $2.4 \pm 0.2$ & 49.2 & & $22.6 \pm 0.2$ & $-47.7 \pm 0.3$ & 57.5 \\ \cline{2-9}
         & BML\_div\tnote{e} & &               &      & & $26.0 \pm 0.3$ & $-44.3 \pm 0.4$ & 363 \\
         & BML\_rot\tnote{f} & &               &      & & $25.7 \pm 0.2$ & $-45.0 \pm 0.3$ & 41.9 \\
         \cline{2-9}
         & MML\tnote{g}      & & $0.8 \pm 0.2$ & 147  & & $24.9 \pm 0.2$ & $-45.7 \pm 0.3$ & 162 \\
         & MML\_div\tnote{h} & &               &      & & $24.8 \pm 0.3$ & $-44.9 \pm 0.4$ & 59.2 \\
         & MML\_rot\tnote{i} & &               &      & & $24.0 \pm 0.2$ & $-44.5 \pm 0.3$ & 30.1 \\
  \hline 
  4.5    & IM       & & $0.6 \pm 0.2$ & 145\tnote{j} & & $42.0 \pm 0.2$ & $-44.6 \pm 0.2$ & 173\tnote{j} \\
         & GR       & & $1.4 \pm 0.2$ & 221\tnote{j} & & $41.0 \pm 0.2$ & $-43.8 \pm 0.2$ & 283\tnote{j} \\
         \cline{2-9}
         & BML\_div & &               &       & & $46.8 \pm 0.3$ & $-45.0 \pm 0.2$ & 347 \\
         & BML\_rot & &               &       & & $47.3 \pm 0.2$ & $-45.2 \pm 0.2$ & 44.2 \\
         \cline{2-9}
         & MML      & & $0.8 \pm 0.2$ & 106   & & $47.6 \pm 0.2$ & $-44.9 \pm 0.1$ & 145 \\
         & MML\_div & &               &       & & $47.0 \pm 0.3$ & $-44.6 \pm 0.2$ & 49.3 \\
         & MML\_rot & &               &       & & $47.5 \pm 0.2$ & $-44.8 \pm 0.1$ & 35.6 \\
  \hline 
  6.4    & IM       & & $0.8 \pm 0.2$ & 45.8 & & $54.3 \pm 0.2$ & $-44.6 \pm 0.1$ & 34.1 \\
         & GR       & & $1.3 \pm 0.2$ & 114  & & $52.6 \pm 0.2$ & $-44.2 \pm 0.1$ & 119 \\
         \cline{2-9}
         & BML\_div & &               &      & & $62.6 \pm 0.3$ & $-45.2 \pm 0.1$ & 403 \\
         & BML\_rot & &               &      & & $62.3 \pm 0.2$ & $-45.0 \pm 0.1$ & 22.5 \\
         \cline{2-9}
         & MML      & & $1.1 \pm 0.2$ & 57.5 & & $62.5 \pm 0.2$ & $-45.3 \pm 0.1$ & 96.3 \\
         & MML\_div & &               &      & & $62.2 \pm 0.3$ & $-44.9 \pm 0.1$ & 29.6 \\
         & MML\_rot & &               &      & & $62.3 \pm 0.2$ & $-45.0 \pm 0.1$ & 18.8 \\
  \hline 
  8.0    & IM       & & $0.9 \pm 0.2$ & 56.8 & & $58.1 \pm 0.2$ & $-44.7 \pm 0.1$ & 39.7 \\
         & GR       & & $0.3 \pm 0.2$ & 81.1 & & $58.9 \pm 0.2$ & $-45.0 \pm 0.1$ & 101 \\
         \cline{2-9}
         & BML\_div & &               &      & & $69.2 \pm 0.2$ & $-45.3 \pm 0.1$ & 278 \\
         & BML\_rot & &               &      & & $69.0 \pm 0.2$ & $-45.1 \pm 0.1$ & 33.2 \\
         \cline{2-9}
         & MML      & & $0.6 \pm 0.2$ & 53.8 & & $68.9 \pm 0.2$ & $-45.2 \pm 0.1$ & 101 \\
         & MML\_div & &               &      & & $68.9 \pm 0.2$ & $-45.0 \pm 0.1$ & 43.7 \\
         & MML\_rot & &               &      & & $69.1 \pm 0.2$ & $-45.0 \pm 0.1$ & 31.3 \\
  \hline 
 \end{tabular}
 \begin{tablenotes}
  \item[a] Perfectly polarized data containing X-rays with a polarization angle of -45~deg.
  \item[b] All degrees of freedom for the $\chi^{2}$ test are 33.
  \item[c] Image moment (IM) method with processing parameters adjusted by actual data \cite{2018NIMPA.880..188K}.
  \item[d] Graph-based (GR) method with parameters adjusted by actual data \cite{2018NIMPA.880..188K}.
  \item[e] Basic machine learning (BML) method with the division approach.
  \item[f] BML method with the polarimeter rotation technique.
  \item[g] Modified machine learning (MML) method
  \item[h] MML method with the division approach for the response unfolding.
  \item[i] MML method with the polarimeter rotation technique.
  \item[j] Relatively poor fit due to difficulty in accurately simulating photoelectron tracks with an energy of $\sim4.5$~keV, where the track range is comparable to the electron diffusion size.
 \end{tablenotes}
\end{threeparttable}

\begin{figure}
 \begin{center}
  \includegraphics[width=7.0cm]{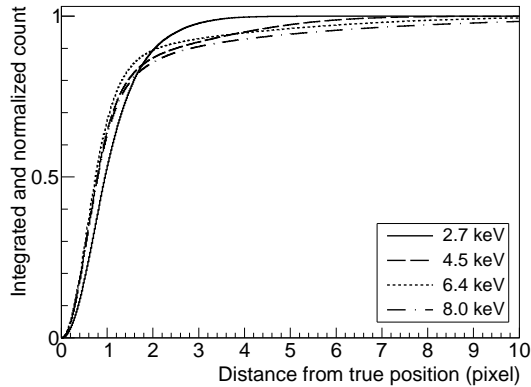}
 \end{center}
 \caption
 { \label{fig:CompHPD}
 Cumulative distribution of the distance between predicted and true positions for different monochromatic X-rays. HPDs for (solid) 2.7, (dashed) 4.5, (dotte) 6.4, and (dash-dotted) 8.0~keV X-rays are 1.9, 1.6, 1.5, and 1.6 pixel size (or 240, 190, 180, and 200~$\mu$m in the actual size), respectively.}
\end{figure}

In addition to the initial direction of the photoelectron, the developed network is designed to simultaneously predict the initial position. A cumulative distribution of the distance between predicted and true positions shown in Fig.~\ref{fig:CompHPD} displays that half power diameters (HPDs) at 2.7, 4.5, 6.4, and 8.0~keV are 1.9, 1.6, 1.5, and 1.6~pixel (or 240, 190, 180, and 200~$\mu$m), respectively. The HPDs decrease with increasing energy except for 8.0~keV. The exception is because some of longer tracks have an interaction position that protrudes outside the captured image frame. The machine-predicted HPDs are improved compared with graph-based ones which are 2.3, 3.4, 3.2, and 3.7~pixel at 2.7, 4.5, 6.4, and 8.0~keV, respectively.

In order to evaluate result reproducibility provided by machine learning, the network model is retrained 10 times in the same way but with different randomized initial values. Although the irregular residual modulation pattern varies with every training, the 8.0~keV modulation factors derived with the uniform rotation method are scattered with the standard deviation of 0.1\% and the average value of 68.9\%. The standard deviation for the other energies are almost the same level. In addition, position accuracy is found to be unchanged. Therefore, the developed network can reproduce the results within 0.1\% for all the energies.

\subsection{Modified model to uniform predictions}
\label{sect:mod}

Although machine learning can improve the modulation factors by 10--20\% as described in \S~\ref{sect:bas}, the unpolarized modulation curve is not flat. The non-flat angular response can potentially cause the dependence of the modulation factor on the polarization angle. According to Ref.~\citenum{2017ApJ...838...72S}, if the polarization-angular response depends on the difference of the polarization angle and the observed angle and thus becomes flat for unpolarized X-rays, the angular response function can be separated into two independent functions. One is the 2-D energy response matrix used widely for spectral fitting. The other is the modulation function which is a 1-d vector as a function of the intrinsic energy and can be processed in the similar way as an effective area (or a so-called ancillary response) function in the XSPEC spectral fitting package \cite{1996ASPC..101...17A}. Therefore, in order to simplify the polarimetry analysis and use the XSPEC heritage, the modulation curve for unpolarized X-rays is preferable to be flat.

In order to obtain a uniform (flat) angular response for unpolarized X-rays, a nonuniform penalty term should be added to the loss function. An additional loss term must have a gradient (or be differentiable) with respect to individual predictions, which are inputs to the operation to calculate a nonuniformity, to perform backpropagation. However, the histogram process to create a modulation curve and evaluate an angular uniformity is not continuously differentiable. Therefore, an unbinned method to evaluate a uniformity is required.

So far, the most probable angle is employed as the angular class with the highest predicted probability. However, this operation is also not continuously differentiable. Instead, the circular mean of the angular probability distribution is used as the predicted angle by calculating as follows:
\begin{equation}
\label{eq:mean}
\bar{\varphi}_{j}(p_{j,i}) = \atantwo \left( \sum^{N_{\mathrm{c}}-1}_{i=0} p_{j,i} \sin \varphi_{i}, ~\sum^{N_{\mathrm{c}}-1}_{i=0} p_{j,i} \cos \varphi_{i} \right),
\end{equation}
where $\bar{\varphi}_{j}$ is the circular mean angle of the $j$-th event in a mini-batch, $N_{\mathrm{c}}$ is the number of angular classes, $p_{j,i}$ is a predicted probability in the $i$-th class for the $j$-th event, and $\varphi_{i}$ is the center angle of the $i$-th class ($\varphi_{i} = \pi(2i+1) / N_{\mathrm{c}}$). It should be noted that $\varphi$ ranges from -180 to 180~deg, while the predicted angle $\phi$ ranges from -90 to 90~deg. This is because the circular mean is valid for angles ranging from -180 to 180~deg. It is also worth noting that Eq.~(\ref{eq:mean}) is differentiable with respect to the input $p_{j,i}$ because it consists of a combination of trigonometric functions. The angle $\bar{\varphi}_{j}$ is a continuous value unlike a discrete angle predicted in the previous section.

As an unbinned method to evaluate a directional uniformity, the $H$-test has been widely used for periodic signal search in X-ray/$\gamma$-ray astronomy \cite{1989A&A...221..180D}. The $H$-test is based on the test statistics $Z^{2}_{m}$ that is the sum of the Fourier powers of the first $m$ harmonics as given by:
\begin{equation}
\label{eq:Z2m}
 Z^{2}_{m}(\bar{\varphi}_{j}) = \frac{2}{N} \sum_{k=1}^{m}
  \left[ \left( \sum_{j=0}^{N-1}\cos k \bar{\varphi}_{j} \right)^{2}
   + \left( \sum_{j=0}^{N-1}\sin k \bar{\varphi}_{j} \right)^{2} \right],
\end{equation}
where $N$ is the number of events considered. The null distribution of $Z^{2}_{m}$ approaches a $\chi^{2}$ distribution with $m\times2$ degrees of freedom as $N$ increases. The gradient of Eq.~(\ref{eq:Z2m}) with respect to $\bar{\varphi}_{j}$ can be easily described as:
\begin{equation}
\label{eq:dZ2m}
 \frac{\partial Z^{2}_{m}}{\partial \bar{\varphi}_{j}}
  = \frac{4}{N} \sum_{k=1}^{m}
  k \left[ - \sin k \bar{\varphi}_{j} \left( \sum_{j=1}^{N} \cos k \bar{\varphi}_{j} \right)
     + \cos k \bar{\varphi}_{j} \left( \sum_{j=1}^{N} \sin k \bar{\varphi}_{j} \right) \right].
\end{equation}
By using $Z^{2}_{m}$, the $H$ statistics, $H_{M}$ described to avoid confusion with the cross entropy $H$, is defined as:
\begin{equation}
\label{eq:H}
 H_{M} = \max_{1 \leq m \leq 20}(Z^{2}_{m} - 4m + 4) = Z^{2}_{M} - 4M + 4 \geq 0,
\end{equation}
where $M$ is $m$ with the maximum of $Z^{2}_{m} - 4m + 4$. The $H$ statistics is non-negative because $H_{M} \geq H_{1} = Z^{2}_1 \geq 0$. Furthermore, the deviation of $H_{M}$ is  $\partial H_{M} / \partial \bar{\varphi}_{j} = \partial Z^{2}_{M} / \partial \bar{\varphi}_{j}$. Therefore, it can be used as the loss function penalizing angular nonuniformity -- learning with backpropagation proceeds by minimizing it.

In the practical observation, spectral polarimetry with energy-resolved modulation curves will be conducted. Therefore, energy-resolved $H_{M}$ values are calculated so that each modulation curve becomes flat. For that reason, the mini-batch data containing 0.5--12~keV X-rays are separated by the incident X-ray energy into 40 ranges, each having $\sim0.3$~keV width. Then, the mean value, $\bar{H}_{M}$, of non-zero elements is calculated along the 40 energy ranges.

A modified loss function by adding the $H$ statistics to the loss fuction in Eq.~(\ref{eq:LH2}) is given by
\begin{equation}
\label{eq:LHZ}
 L = \sum H + \lambda_{2} \sum w^{2} + \lambda_{H} \bar{H}_{M},
\end{equation}
where the first and third terms respectively represent the classification performance and the angular uniformity. $\lambda_{H}$ is a hyperparameter for the $H$ statistics to control the trade-off between the two qualities and set to be 0.01. The hyperparameter $\lambda_{2}$ is set to be the same value as in the previous learning, 0.007, because the number of weights making up the network model is unchanged.

Machine learning is performed with the same manner and same training and verification data sets as described in \S~\ref{sect:bas}, but with consideration of the uniform angular predictions by replacing the loss function with Eq.~(\ref{eq:LHZ}) and with a larger mini-batch size. The mini-batch size should be large enough to reduce the statistical error of uniformity and is therefore set to be 18,000 limited by the GPU memory size of 8~GB. Additionally, the training images with the total size of 3,600,000 are sorted by irradiation position (or electron drift length) and then are grouped into 200 mini-batches. This is because the predicted modulation curve should be flat along the irradiation position as well as the incident X-ray energy. Although energy- and position-sorted images are inputted, learning efficiency becomes low due to biased track shapes in a mini-batch. Instead, uniformity along energies is evaluated with energy-resolved $H_{M}$ in Eq.~(\ref{eq:LHZ}).

\begin{figure}
 \begin{center}
  \includegraphics[width=7.0cm]{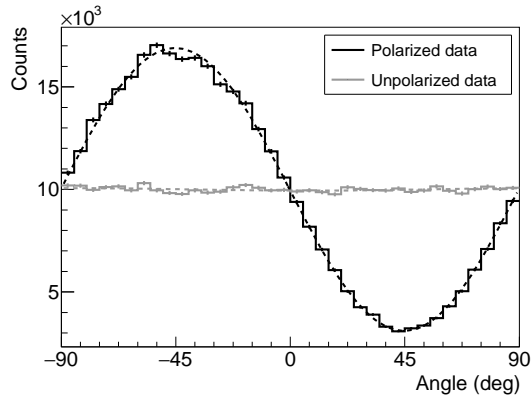}
 \end{center}
 \caption
 { \label{fig:UnifMC}
 Same as Fig.~\ref{fig:UnfoldMC}~(a), but with the modified model with nonuniform penalty. The dashed curves for polarized and unpolarized data show the best-fit sinusoidal model with a constant offset, each having a modulation amplitude of 68.9 and 0.6\%, respectively.}
\end{figure}

In the same way as described in \S~\ref{sect:bas}, the network model is learned, evaluated after finishing every epoch, saved at the lowest loss, and tested by inputting the unpolarized and polarized monoenergetic test data sets. The machine learning lasts for 1500~epochs or $\sim 110$ hours because learning efficiency is lower than that in the previous model. The 8.0~keV modulation curves in Fig.~\ref{fig:UnifMC}, which are derived from predicted angles $\bar{\phi} = \bar{\varphi}/2$ with the circular mean in Eq.~(\ref{eq:mean}), clearly illustrate that the residual modulation in the unpolarized data is much suppressed. In addition, the polarized modulation is close to a unimodal sinusoid. However, the residual modulation still remains indicated by a reduced $\chi^{2}$ value of $\sim 3$, which is derived by fitting the sinusoidal curve model to the modulation curve. The fitting results including those for the other energies are listed in Table~\ref{tab:summary}.

The residual systematic modulation can be canceled out by the division or rotation method as already described in \S~\ref{sect:bas}. The best-fit parameters with the two methods are also listed in Table~\ref{tab:summary}. It shows that both models can improve $\chi^{2}$. In particular, the divided modulation curves can be reproduced by the sinusoidal curve model with reasonable $\chi^{2}$ because it is empirically known that the division method becomes better approximation for response unfolding as the denominator histogram is more featureless. The modulation factors are degraded only by $<2$\% as an absolute difference, compared to those predicted by the basic network model. The best-fit polarization degrees at the four energies are scattered around the set polarization angle of -45~deg within $<1$~deg. In addition, it is also confirmed that the determination accuracy of the initial photoelectron position is unchanged.

\subsection{Track selection to improve sensitivity}
\label{sect:sel}

Reduction of tracks, from which it is difficult to extract the initial photoelectron angle, can improve the polarization sensitivity. So far, track eccentricity and circularity were demonstrated to quantify track roundness and discard round tracks (e.g. Refs.~\citenum{2010NIMPA.620..285M,2012SPIE.8443E..4KB,2016NIMPA.838...89I,2018NIMPA.880..188K}). However, it is complicated to optimize a cut region in the two (or more) parameter plane (or space). Instead, the maximum value of the angular probability distribution, $p_{\mathrm{max}}$, for each track outputted from the network model is suggested.

\begin{figure}
 \begin{center}
  \includegraphics[width=7.0cm]{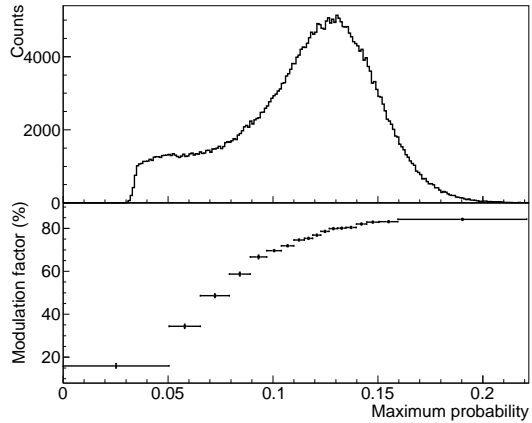}
 \end{center}
 \caption
 { \label{fig:MaxProb}
 (Top) Histogram of the maximum probability, $p_{\mathrm{max}}$, of angular predictions for each photoelectron track generated by 8.0~keV X-rays and (bottom) the modulation factor at 8.0~keV as a function of $p_{\mathrm{max}}$. Each modulation factor in the bottom panel is derived from the modulation curve consisting of $\sim20,000$ tracks with the probability range indicated by the horizontal line of each data point and is calculated with the polarimeter rotation method in \S~\ref{sect:bas}.}
\end{figure}

The test data sets, which were processed with the modified network model at the minimum loss and with uniform rotation along the optical axis in \S~\ref{sect:bas}, are used to demonstrate track selection with $p_{\mathrm{max}}$. As shown in Fig.~\ref{fig:MaxProb}, the modulation factors derived from modulation curves sorted by $p_{\mathrm{max}}$ monotonically increase with $p_{\mathrm{max}}$. Therefore, we set a threshold for $p_{\mathrm{max}}$ to separate valid from invalid tracks and discard tracks with lower $p_{\mathrm{max}}$.

\begin{figure}
 \begin{center}
  \includegraphics[width=7.0cm]{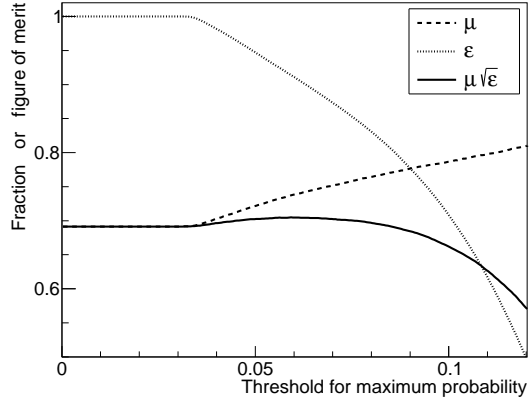}
 \end{center}
 \caption
 { \label{fig:FoM}
 Figure of merit curve to maximize the polarimetry sensitivity for 8.0~keV X-rays as a function of threshold for the maximum probability $p_{\mathrm{max}}$. The dashed and dotted curves are respectively the modulation factor $\mu$ and signal acceptance $\varepsilon$ which are derived from images with the maximum probability $p_{\mathrm{max}}$ above the set threshold. The solid curve represent the figure of merit $F \equiv \mu \sqrt{\varepsilon}$, where the peak at a threshold of 0.0593 indicates the maximum polarimeter sensitivity.}
\end{figure}

In order to evaluate and optimize track selection with $p_{\mathrm{max}}$, the figure of merit, $F$, is defined in the same way as Ref.~\citenum{2018NIMPA.880..188K} as given by $F \equiv \mu \sqrt{\varepsilon}$, where $\varepsilon$ is signal acceptance decreased by track selection and $\mu$ is the modulation factor of selected events. As shown in Fig.~\ref{fig:FoM}, $\varepsilon$ monotonically decreases with increasing the threshold for $p_{\mathrm{max}}$, while $\mu$ monotonically increases. We search for the best selection condition where $F$ for 8.0~keV X-rays is maximized by discarding tracks in order of increasing $p_{\mathrm{max}}$, and find that $F$ with $p_{\mathrm{max}} > 0.0593$, where $\varepsilon = 91.3$\% and $\mu=73.8$\%, is improved by 2.0\% compared to $F$ with no track selection. On the other hand, conventional track selection with eccentricity derived from the corresponding processed image can also improve $F$ by 0.03\%. The results for the other test energies are obtained in the same manner and listed in Table~\ref{tab:sens}. It shows that $p_{\mathrm{max}}$ is a better selector than eccentricity for all the energies.

Table~\ref{tab:sens} also lists reduction rates of the observation time to achieve a given polarimeter sensitivity, which increases with the square root of the observation time if the background rate is negligible. The machine-learning-based method for angular reconstruction and track selection can reduce the observation time by $\sim$20--30\%, compared to the conventional IM method.

\begin{threeparttable}
 \centering
 \caption
 { \label{tab:sens}
 Comparison of polarimeter sensitivity derived from the various angular reconstruction methods combined with track selection.}
 \begin{tabular}{l l l l l l l l l}
  \hline 
  \hline 
  Energy & Method & \multicolumn{1}{c}{$\mu$} & \multicolumn{1}{c}{$\phi_{0}$} & \multicolumn{1}{c}{$\chi^{2}$~\tnote{a}} & \multicolumn{1}{c}{$t$\tnote{b}} & \multicolumn{1}{c}{$\varepsilon$} & \multicolumn{1}{c}{$F$\tnote{c}} &  \multicolumn{1}{c}{$R_{T}$\tnote{d}} \\
  \multicolumn{1}{c}{(keV)} & & \multicolumn{1}{c}{(\%)} & \multicolumn{1}{c}{(deg)} & & & & & \\
  \hline 
  2.7    & IM\tnote{e}             & $23.4 \pm 0.2$ & $-45.1 \pm 0.3$ & 41.4 & 0      & 1     & 0.234 & 1 \\
         & IM\_ecut\tnote{f}       & $27.7 \pm 0.3$ & $-45.2 \pm 0.3$ & 33.0 & 0.490  & 0.766 & 0.243 & 0.927 \\
         & MML\_rot\tnote{g}       & $24.0 \pm 0.2$ & $-44.5 \pm 0.3$ & 30.1 & 0      & 1     & 0.240 & 0.951 \\
         & MML\_rot\_ecut\tnote{h} & $26.2 \pm 0.2$ & $-44.6 \pm 0.3$ & 29.2 & 0.431  & 0.860 & 0.242 & 0.935 \\
         & MML\_rot\_pcut\tnote{i} & $30.5 \pm 0.3$ & $-44.8 \pm 0.3$ & 31.0 & 0.0388 & 0.708 & 0.257 & 0.829 \\
  \hline 
  4.5    & IM                      & $42.0 \pm 0.2$ & $-44.6 \pm 0.2$ & 173  & 0      & 1     & 0.420 & 1 \\
         & IM\_ecut                & $45.1 \pm 0.2$ & $-44.6 \pm 0.2$ & 181  & 0.550  & 0.896 & 0.427 & 0.967 \\
         & MML\_rot                & $47.5 \pm 0.2$ & $-44.8 \pm 0.1$ & 35.6 & 0      & 1     & 0.475 & 0.782 \\
         & MML\_rot\_ecut          & $50.2 \pm 0.2$ & $-44.7 \pm 0.1$ & 36.6 & 0.538  & 0.908 & 0.478 & 0.772 \\
         & MML\_rot\_pcut          & $53.4 \pm 0.2$ & $-44.8 \pm 0.1$ & 38.7 & 0.0488 & 0.834 & 0.488 & 0.741 \\
  \hline 
  6.4    & IM                      & $54.3 \pm 0.2$ & $-44.6 \pm 0.1$ & 34.1 & 0      & 1     & 0.543 & 1 \\
         & IM\_ecut                & $57.7 \pm 0.2$ & $-44.5 \pm 0.1$ & 30.7 & 0.705  & 0.908 & 0.550 & 0.975 \\
         & MML\_rot                & $62.3 \pm 0.2$ & $-45.0 \pm 0.1$ & 18.8 & 0      & 1     & 0.623 & 0.760 \\
         & MML\_rot\_ecut          & $62.6 \pm 0.2$ & $-45.0 \pm 0.1$ & 18.8 & 0.398  & 0.993 & 0.623 & 0.760 \\
         & MML\_rot\_pcut          & $68.6 \pm 0.2$ & $-45.0 \pm 0.1$ & 21.3 & 0.0563 & 0.867 & 0.639 & 0.722 \\
  \hline 
  8.0    & IM                      & $58.1 \pm 0.2$ & $-44.7 \pm 0.1$ & 39.7 & 0      & 1     & 0.581 & 1 \\
         & IM\_ecut                & $60.2 \pm 0.2$ & $-44.7 \pm 0.1$ & 43.7 & 0.710  & 0.946 & 0.585 & 0.986 \\
         & MML\_rot                & $69.1 \pm 0.2$ & $-45.0 \pm 0.1$ & 31.3 & 0      & 1     & 0.691 & 0.707 \\
         & MML\_rot\_ecut          & $69.2 \pm 0.2$ & $-45.0 \pm 0.1$ & 31.6 & 0.325  & 0.998 & 0.692 & 0.705 \\
         & MML\_rot\_pcut          & $73.8 \pm 0.2$ & $-45.0 \pm 0.1$ & 36.4 & 0.0593 & 0.913 & 0.705 & 0.679 \\
  \hline 
 \end{tabular}
 \begin{tablenotes}
  \item[a] All degrees of freedom for the $\chi^{2}$ test are 33.
  \item[b] Threshold $t$ of eccentricity for the IM method and the maximum probability $p_\mathrm{max}$ for the MML method. Tracks below the threshold are discarded.
  \item[c] Figure of merit, $F \equiv \mu \sqrt{\varepsilon}$ (higher is better), representing the polarimeter sensitivity.
  \item[d] Reduction rate of the observation time to achieve a given polarimetry sensitivity, compared to the IM method.
  \item[e] Same as the IM method in Table~\ref{tab:summary}
  \item[f] IM method with track selection by using eccentricity.
  \item[g] Same as the MML\_rot method in Table~\ref{tab:summary}.
  \item[h] MML\_rot method with track selection by using eccentricity.
  \item[i] MML\_rot method with track selection by using $p_\mathrm{max}$.
 \end{tablenotes}
\end{threeparttable}

\section{Summary and Discussion}
\label{sect:con}

We have developed angular and positional reconstruction of a photoelectron track image for the TPC X-ray polarimeter by using supervised machine learning. Data sets for training, validation, and test are generated by the Monte Carlo simulation we have developed. We test two CNN classifications; the VGG-based model and the modified one with a penalty term for nonuniformity of angular predictions. Although the former network outputs large residual modulation for unpolarized X-rays (see Fig.~\ref{fig:CompMC}), it can be canceled out by unfolding the response or rotating the polarimeter. The resulting modulation factor is improved by $\sim$10--20\% for all the energies, compared to the conventional reconstruction methods. The latter network considers uniformity of angular predictions, and therefore residual modulation for unpolarized X-rays is drastically reduced with decreasing the modulation factor within 2\% (see Fig.~\ref{fig:UnifMC} and Table~\ref{tab:summary}). In addition, we suggest track selection with the maximum probability of predictions to further improve the polarimeter sensitivity.

The modified network with the nonuniform penalty still shows the residual modulation, as indicated by the relatively large $\chi^{2}$ (for the MML method in Table~\ref{tab:summary}). The residual modulation may be reduced by increasing the mini-batch size, though it is limited by the GPU memory size. Therefore, a GPU with a larger memory size is preferable. Alternatively, a more recent network with smaller memory consumption but deeper layers based on e.g. Inception \cite{2016arXiv160207261S} and/or ResNet \cite{2015arXiv151203385H} architecture could increase the mini-batch size as well as improve the modulation factor (see Ref.~\citenum{2016arXiv160507678C} for information on network architecture comparisons).

The developed method would be applicable to not only the other photoelectric polarimeters, such as the Gas Pixel Detector (GPD) polarimeter \cite{2001Natur.411..662C,2007NIMPA.579..853B}, but also electron-tracking Compton polarimeters which are capable of determining the initial emission direction of recoil electrons to improve Compton kinematics reconstruction (e.g. \cite{2017NatSR...741511T,2018NIMPA.912..269Y}). However, the GPD polarimeter takes track images in a hexagonal grid to reduce residual modulation, and therefore requires a hexagonal convolution operation. A simple way demonstrated in Ref.~\citenum{2019APh...105...44S}, where the CNN is applied to H.E.S.S. hexagonal images, is that the hexagonal convolution can be performed with a combination of available operations for the square convolution.

\section{Future work}

Observed track images can be input to the network model trained by the simulated images to extract polarization information. However, the residual modulation is most likely to appear due to systematic effects induced by a real polarimeter and an imperfect simulation. Although the residual modulation can be canceled out with the division method or the rotation method, the modulation factor is probably degraded compared to that derived from the simulated data. Training with observed rather than simulated data should be better for practical polarimetry observation.

The developed method needs the initial emission angle of each photoelectron track, and therefore demands as accurate a simulator as possible to generate a realistic image with the known angle. However, the network model can be potentially supervised with the modulation factor instead of the angles. Our method calculates directional uniformity, which is a statistic of samples, and uses it for training of machine learning. The modulation factor is also a sample statistic and can be derived from the Stokes parameters on an event-by-event basis \cite{2015APh....68...45K} without using a histogram to create a modulation curve. The modulation factor, $\mu$, can be alternatively represented by $\mu = 2 \sqrt{ \left( \sum^{N-1}_{i=0} \cos \bar{\varphi_{i}} \right)^{2} + \left( ~\sum^{N-1}_{i=0} \sin \bar{\varphi_{i}} \right)^{2}} / N = \sqrt{2 Z^{2}_{1} / N}$, which ranges from 0 to 1. The polarization angle, $\varphi_{0}$, can also be given by $\varphi_{0} = \atantwo \left( \sum^{N-1}_{i=0} \sin \bar{\varphi_{i}}, ~\sum^{N-1}_{i=0} \cos \bar{\varphi_{i}} \right)/2$. These operations are differentiable and therefore are available for backpropagation to supervise machine learning. The cross entropy $H$ in Eq.~(\ref{eq:LHZ}) can be replaced with the two statistical values to represent the learning score, meaning that the revised method can input an observed data set with known Stokes parameters (or the polarization degree and angle) of incident X-rays -- then a simulator is no longer needed because the initial angle of each image is not needed for training. The above revised method will be demonstrated and discussed elsewhere.

\section*{Acknowledgments}
The authors would like to acknowledge helpful discussion with Keith Jahoda on polarimeter development.
This work was partially supported by JSPS KAKENHI Grant Numbers JP17K18776 and JP16H02198.

\end{document}